\documentclass[12pt]{article}

\usepackage[small]{caption}
\usepackage{color}
\definecolor{darkblue}{rgb}{0.1,0.1,.7}
\usepackage[dvips, colorlinks, linkcolor=darkblue, citecolor=darkblue, urlcolor=black, linktocpage]{hyperref}
\usepackage{epsfig,amsmath,amssymb,verbatim,mathrsfs}
\usepackage[square, comma, sort&compress,numbers]{natbib}
\usepackage{dsdshorthand}
\usepackage{feynmp}
\usepackage{subfigure}

\numberwithin{equation}{section}
\newcommand\cN{\mathcal{N}}

\newcommand\lrptl{\raise .8ex\hbox{$^\leftrightarrow$} \hspace{-9pt}\partial}
\newcommand\nn{\nonumber}
\newcommand\up{\uparrow}
\newcommand\down{\downarrow}
\newcommand\id{\mathbb{I}}

\renewcommand\be{\begin{eqnarray}}
\renewcommand\ee{\end{eqnarray}}

\newcommand\biblink[2]{{\tt\href{#1}{#2}}}

\title{{\Large {\bf $\cN=1$ SQCD and the Transverse Field Ising Model}}}
\author{David Poland and David Simmons-Duffin \\ \\
{\it \normalsize Jefferson Physical Laboratory, Harvard University,}\\
{\it \normalsize Cambridge, Massachusetts 02138, USA}}

\begin{document}

\begin{titlepage}

\noindent

\vspace{1cm}

\maketitle
\thispagestyle{empty}

\begin{abstract}
We study the dimensions of non-chiral operators in the Veneziano limit of $\cN=1$ supersymmetric QCD in the conformal window.  We show that when acting on gauge-invariant operators built out of scalars, the 1-loop dilatation operator is equivalent to the spin chain Hamiltonian of the 1D Ising model in a transverse magnetic field, which is a nontrivial integrable system that is exactly solvable at finite length.  Solutions with periodic boundary conditions give the anomalous dimensions of flavor-singlet operators and solutions with fixed boundary conditions give the anomalous dimensions of operators whose ends contain open flavor indices.
\end{abstract}

\end{titlepage}


\newpage

\section{Introduction}
\label{sec:intro}

Conformal field theories in four dimensions are both ubiquitous and have many potential phenomenological applications (e.g.,~\cite{Georgi:1983mq,Holdom:1984sk,Akiba:1985rr,Appelquist:1986an,Yamawaki:1985zg,Appelquist:1986tr,Appelquist:1987fc,ArkaniHamed:2000ds,Nelson:2000sn,Kobayashi:2001kz,Nelson:2001mq,Luty:2001jh, Luty:2001zv, Kobayashi:2002iz,Dine:2004dv,Luty:2004ye,Sundrum:2004un,Ibe:2005pj, Ibe:2005qv,Nomura:2006pn,Schmaltz:2006qs,Roy:2007nz, Murayama:2007ge,Luty:2008vs,Poland:2009yb,Galloway:2010bp,Baumann:2010ys,Craig:2010ip,Evans:2010ed}). 
While much progress has been made towards unraveling the structure of theories with lots of supersymmetry such as $\cN=4$ SYM, strongly-coupled CFTs with $\cN=1$ or no supersymmetry remain fairly elusive.  In particular, phenomenological applications often depend crucially on the dimensions of unprotected operators, and very few tools are generally available to learn about their dimensions.  While some progress has been made recently towards placing general bounds on unprotected quantities~\cite{Rattazzi:2008pe,Caracciolo:2009bx,Rychkov:2009ij,Poland:2010wg,Rattazzi:2010gj,Rattazzi:2010yc}, the situation is still unsatisfactory and it is important to develop new techniques to learn about 4D CFTs with little or no supersymmetry.

In order to gain a better understanding of these theories it is very useful to find limits in which simplifying structures emerge.  One such limit is that of weak coupling $\l \ll 1$ (possibly in a dual frame), where perturbation theory is valid.  Another is a large-$N$ limit~\cite{'tHooft:1973jz}, where the theory becomes planar, and operator dimensions and correlation functions factorize according to large-$N$ counting.  In the present work, we will focus on $\cN=1$ supersymmetric QCD with gauge group $\SU(N_c)$ in the conformal window $\frac{3}{2} N_c < N_f < 3 N_c$~\cite{Seiberg:1994pq}.  This theory has a large-$N$ limit in the sense of Veneziano~\cite{Veneziano:1976wm}, taking $N_c\to\oo$ while holding $N_f/N_c$ fixed.  In the Veneziano limit it is additionally possible to take $N_f/N_c$ arbitrarily close to either edge of the conformal window, where either the electric or magnetic description of the theory has a perturbative Banks-Zaks fixed point~\cite{Banks:1982} and is weakly coupled.

The remarkable lesson of the AdS/CFT correspondence~\cite{Maldacena:1997re,Gubser:1998bc,Witten:1998qj} is that large-$N$ CFTs at {\it very strong} coupling $\l \gg 1$ can be described in terms of perturbative string theory in AdS backgrounds.  Moreover, the low-dimension spectrum of such theories can often be accurately described by effective field theories in AdS.  Unfortunately, this limit is typically only accessible in theories with an exactly marginal coupling, such as $\cN=4$ SYM.  By contrast $\cN=1$ SQCD flows to an isolated fixed point, where the 't Hooft coupling is generically expected to be $O(1)$.  A quantitative string-theoretic description of these fixed points is currently lacking, and any such description is generally not expected to admit a traditional effective field theory limit~\cite{Klebanov:2004ya,Bigazzi:2005md}.

The last decade has also seen remarkable progress in understanding certain special theories (including $\cN=4$ SYM and some deformations) at {\it arbitrary} values of the 't Hooft coupling.  This is due to the property of integrability, which roughly means that the theory has an infinite number of hidden symmetries not manifest in the Lagrangian description (see~\cite{Beisert:2010jr} for a review and many references).  Some of the first hints for integrability in $\cN=4$ SYM came from calculations of the action of the 1-loop dilatation operator on closed sectors of operators built out of scalar fields.  In~\cite{Minahan:2002ve} it was shown that the 1-loop dilatation operator acting on the $\SO(6)$ subsector of scalar operators (closed at 1-loop) is identical to the Hamiltonian for an integrable $\SO(6)$ Heisenberg spin chain, which may be diagonalized using a Bethe ansatz.  This investigation paved the way to discovering integrability of the full 1-loop dilatation operator of the theory~\cite{Beisert:2003tq,Beisert:2003yb,Beisert:2003jj,Beisert:2004ry}, and was soon generalized to various sectors of the gauge theory at the higher-loop level (e.g.,~\cite{Beisert:2003jb,Beisert:2003ys,Serban:2004jf,Ryzhov:2004nz,Beisert:2004hm,Minahan:2004ds,Staudacher:2004tk,Beisert:2005fw,Beisert:2005tm,Beisert:2006qh,Beisert:2006ez,Beisert:2007hz}).

While it would be quite surprising to find higher-loop integrability in a wide range of theories, the success of the above program suggests that potentially interesting structures might lie hidden in other 4D CFTs.  Moreover, it is not yet clear precisely what role supersymmetry plays in creating such structures.  We believe that it is important to gather more data about how and when simplifying structures can emerge in theories qualitatively different from $\cN=4$ SYM, particularly those with potential phenomenological relevance.

Thus, in the present work we undertake a similar investigation to that of \cite{Minahan:2002ve} for the Veneziano limit of $\cN=1$ SQCD.  We will show that when acting on gauge-invariant operators built out of scalars, the 1-loop dilatation operator (in the electric Banks-Zaks limit, $N_f \sim 3 N_c$) is identical to the Hamiltonian for the 1-dimensional Ising model in a transverse magnetic field.  Here each site on the spin chain is occupied by a flavor-contracted (gauge adjoint) ``dimer", either $Q Q^{\dagger}$ or $\tl{Q}^{\dagger} \tl{Q}$.  This is perhaps one of the simplest examples of a nontrivial integrable model, and can be mapped to a system of quasi-free fermionic excitations through a combination of Jordan-Wigner and Bogoliubov transformations.  We exhibit the exact solutions at arbitrary finite length for both periodic and fixed boundary conditions, which give the 1-loop anomalous dimensions of operators containing both closed and open chains of scalar quarks.

An important difference between this sector of operators and the sectors typically studied in $\cN=4$ SYM is that the ground state is non-BPS, and receives corrections at each order in perturbation theory.  
Indeed, simply identifying the 1-loop ground state in terms of elementary fields is a nontrivial task.  By contrast, in $\cN=4$ SYM one can consider sectors of operators where the ground state is BPS and its form and dimension is protected to all orders. In such cases, the superalgebra preserving the ground state acts linearly on excitations above it~\cite{Berenstein:2002jq}, which is helpful in solving the dynamics and making all-loops predictions for the dimensions of operators~\cite{Beisert:2005fw, Beisert:2005tm,Gromov:2009tv,Bombardelli:2009ns,Gromov:2009bc,Gromov:2009zb}.  We have not found a similar picture in $\cN=1$ SQCD, and it is not yet clear how one would make precise conjectures about the spectrum deeper into the conformal window even if integrability were to persist.

Another notable difference is that excitations above the ground state in $\cN=1$ SQCD are highly nonlocal collective modes, in contrast to the local magnon excitations typically studied in $\cN=4$ SYM.  We expect that similar nonlocal excitations might be present in other theories with a Veneziano limit, such as $\cN=2$ SQCD with gauge group $\SU(N_c)$ and $N_f = 2 N_c$ flavors.  The 1-loop dilatation operator of this theory was recently explored in~\cite{Gadde:2009dj,Gadde:2010zi}, with intriguing but inconclusive results about its integrability properties.  We hope that the present study might be useful in better understanding the structures that appear in this and related theories.

This paper is organized as follows.  In section~\ref{sec:Veneziano} we discuss the Veneziano limit of $\SU(N_c)$ SQCD in the conformal window.  In section~\ref{sec:dilatation} we derive the form of the 1-loop dilatation operator acting on operators built out of scalar fields, and show that it is equivalent to the Hamiltonian of the 1D transverse field Ising model.  In section~\ref{sec:solution} we review the exact solution to this system at arbitrary length for both periodic and fixed boundary conditions, and discuss the resulting spectrum of operator dimensions.  We conclude in section~\ref{sec:concl}.

\section{SQCD in the Veneziano Limit} 
\label{sec:Veneziano}

We will be primarily interested in $\cN=1$ SQCD with gauge group $\SU(N_c)$ and $N_f$ flavors of quarks, $Q_{a i}$ and $\tl{Q}^{\tl{\imath}a}$ ($i,\tl{\imath}=1 \dots N_f$, $a=1 \dots N_c$).  In the range $\frac{3}{2} N_c < N_f < 3 N_c$ the theory is believed to flow to an interacting conformal fixed point~\cite{Seiberg:1994pq}.  The theory has anomaly-free global symmetries $SU(N_f)_L \times SU(N_f)_R \times U(1)_B \times U(1)_R$, where $Q$ transforms as $\left(N_f, 1, 1, 1-N_c/N_f \right)$ and $\tl{Q}$ transforms as $\left(1, \bar{N}_f, -1, 1 - N_c/N_f \right)$.

The Veneziano limit~\cite{Veneziano:1976wm} corresponds to taking $N_c \rightarrow \infty$ while holding $N_f/N_c$ fixed.  In this limit the theory contains a single continuous parameter $\e \equiv 3N_c/N_f - 1$, and when $\e \ll 1$, the electric description is weakly coupled and one can reliably perform calculations in perturbation theory~\cite{Banks:1982}.  The fixed-point value of the 't Hooft coupling then has the expansion~\cite{Gardi:1998ch}\footnote{Here we give the expansion for the coupling as defined in the DRED renormalization scheme~\cite{DRED}.}
\ben
\l \equiv \frac{g^2 N_c}{8\pi^2} = \e + \frac{1}{2} \e^2 + \frac{9}{4} (1+2 \zeta_3) \e^3 + \dots.
\een
One can also compute the dimensions of gauge-invariant operators order by order in $\e$.  A simple example is the chiral meson operator $(\tl{Q} Q)_{i}^{\tl{\imath}}$, whose exact dimension is controlled by the superconformal $U(1)_R$ symmetry.  In this case we have $\De_{\tl{Q} Q} = \frac{3}{2} R_{\tl{Q} Q} = 2 - \e$.  Note that the baryon operators decouple in the large $N_c$ limit, and will play no role in our discussion.

An important simplification that occurs in the $N_c, N_f \rightarrow \infty$ limit is that computations in perturbation theory are dominated by planar diagrams.  Since $N_c$ and $N_f$ are on the same footing when it comes to large-$N$ counting, one must keep track of both color lines and flavor lines in order to decide if a diagram is planar.  As a consequence, the usual large-$N$ factorization story is modified.  In the Veneziano limit, operators factorize into products of ``generalized single-traces," namely strings of fields that have adjacent color {\it and flavor} indices contracted (up to a possible overall trace).  Focusing on operators built only out of the scalars $Q$ and $\tl{Q}$ for simplicity, the basic building blocks of generalized single-trace operators are the flavor-contracted color adjoints $X \equiv Q Q^{\dagger}$ and $Y \equiv \tl{Q}^{\dagger} \tl{Q}$.  From these, one can construct the flavor-singlet gauge-invariant operators
\ben\label{eq:singletrace}
\Tr (X Y\dots X)&=&Q_{ai}Q^{\dag ib}\tl Q^\dag_{b\tl \jmath}\tl Q^{\tl \jmath c}\dots Q_{dk}Q^{\dag ka},
\een
as well as the flavor adjoint and bi-fundamental gauge-invariants
\ben\label{eq:opensingletrace}
(Q^{\dagger} X Y \dots X Q)^i_j,\quad
(Q^{\dagger} X Y \dots X \tl{Q}^{\dagger})^i_{\tl{\jmath}},\quad
(\tl{Q} X Y\dots X Q)^{\tl{\imath}}_j,\quad
(\tl{Q} X Y \dots X \tl{Q}^{\dagger})^{\tl{\imath}}_{\tl{\jmath}}.
\een
Other operators built only out of scalars can in general be written as a product of the operators in Eqs.~(\ref{eq:singletrace}) and (\ref{eq:opensingletrace}), and will be referred to as ``generalized multi-trace" operators.  They have dimensions that are equal to the sum of the dimensions of the constituent single-traces.

Note that operators that are constructed by contracting left flavor indices with right flavor indices,
such as a chain of chiral mesons $\tl Q^{i a} Q_{aj}\tl Q^{j b} Q_{bk}\dots \tl Q^{lc}Q_{ci}$, are generalized {\it multi-}trace operators.  Their correlation functions and dimensions factorize into those of their constituent generalized single-trace parts.\footnote{This can be verified directly by drawing Feynman diagrams with double-line notation as in \cite{Gadde:2009dj}.  With only the interactions of $\cN=1$ SQCD, it is impossible to ``fill in" a region of a diagram between left and right flavor lines.}  If we further insert an ``impurity" into the chain, for example by replacing one of the mesons with $Q^\dag Q$, then the only nontrivial mixing in the anomalous dimension matrix will occur within the generalized single-trace operator that contains the impurity, $\tl Q X Q$.\footnote{In other words, the dispersion relation for such an impurity will be independent of momentum.}  Consequently, it is unclear how to realize in a nontrivial way the $\cN=4$ picture of a BPS ground state sprinkled with impurities, at least using only scalar quarks.

The operators in Eqs.~(\ref{eq:singletrace}) and (\ref{eq:opensingletrace}) have classical dimensions that are equal to the number of constituent quark fields.  At the quantum level, operators carrying the same global symmetry charges can in general mix with each other, and the eigenvectors of this mixing acquire anomalous dimensions that are functions of $\e$.  At the 1-loop level (leading order in $\l \simeq \e$) the above operators mix only with each other and form a closed subsector.  At higher loops they can also mix with operators containing fermions, gauge fields, and derivatives.  In the following sections we will focus on understanding the dimensions of these operators at the 1-loop level.

\section{The 1-Loop Dilatation Operator}
\label{sec:dilatation}

\subsection{Feynman Graphs}

Let us first briefly recall how one computes the dilatation operator in perturbation theory.  Consider a collection of bare operators $\cO^{(0)}_i$ with two-point functions that are canonically normalized at zero coupling $\<\cO^{(0)}_i(x)\cO^{\dag(0)}_j(0)\>_{\l=0}=\de_{ij}x^{-2\De^{(0)}_i}$.  For us, these will be the generalized single-trace operators of Eqs. (\ref{eq:singletrace}) and (\ref{eq:opensingletrace}).  At nonzero coupling, their two-point functions have the form
\be
\label{eq:bare-two-ptfn}
x^{2\De^{(0)}_i}\<\cO^{(0)}_i(x)\cO^{\dag(0)}_j(0)\> &=& \de_{ij}-\l V_{ij}\p{\frac 1 \e +\log(\mu^2 x^2)+\dots}+O(\l^2),
\ee
where we have regulated divergences with dimensional regularization.\footnote{The reader should not confuse the dimensional regularization parameter $\e$ with the expansion parameter $\e=3N_c/N_f-1$ defined in the previous section.}  Now, renormalized operators $\cO_i$ with well-defined scaling dimensions $\De_i$ are linear combinations of bare operators $\cO_i=Z_i{}^j\cO_j^{(0)}$.  The $\De_i$ have a perturbative expansion of the form $\De_i=\De_i^{(0)}+\l\g_i+O(\l^2)$, so that two-point functions of renormalized operators can be written
\be
x^{2\De^{(0)}_i}\<\cO_i(x) \cO_j^\dag(0)\> &=& \de_{ij}x^{-2(\De_i-\De_i^{(0)})}\nn\\
&=& \de_{ij}-\l \g_i \de_{ij} \log(x^2)+O(\l^2)\nn\\
&=&(Z Z^\dag)_{ij}-\l (ZVZ^\dag)_{ij}\p{\frac 1 \e+\log(\mu^2 x^2)}+O(\l^2),
\ee
where in the last line we have substituted Eq.~(\ref{eq:bare-two-ptfn}).  Matching the $O(\l^0)$ term and $1/\e$ pole above, we find $Z=U-\frac{\l}{2\e}UV+\dots$, where $U$ is a unitary matrix, and ``$\dots$" represents scheme-dependent finite terms and general terms of higher order in $\l$.  Finally, matching the coefficients of $\l\log(x^2)$ gives $\g=U V U^\dag$, so that the 1-loop anomalous dimensions are simply the eigenvalues of $V$.

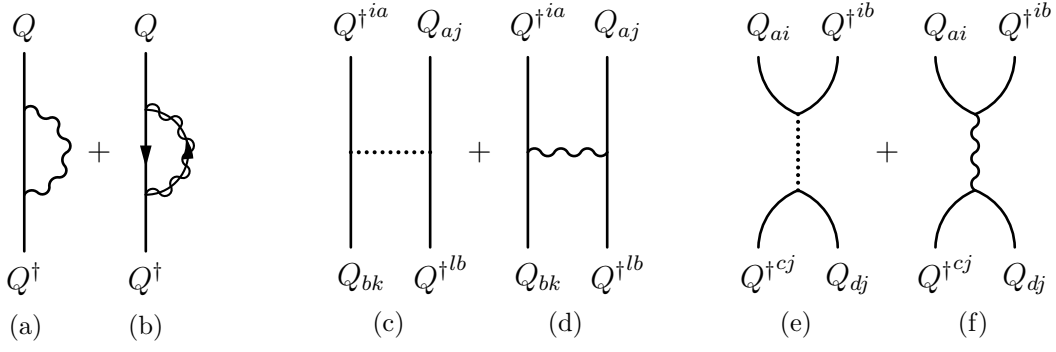
\begin{figure}
\centering
\subfigcapskip=8mm
\subfigure[]{\parbox{4mm}{
\label{fig:wfgraph1}
\begin{fmffile}{wfgraph1}
\begin{fmfgraph*}(30,75)
\fmfset{arrow_len}{3mm}
\fmftop{t}
\fmfbottom{b}
\fmfv{l=$Q$,l.a=90,l.d=1.5mm}{t}
\fmfv{l=$Q^\dag$,l.a=-90,l.d=1.5mm}{b}
\fmf{plain,tension=1.5}{t,v1}
\fmf{plain}{v1,v2}
\fmf{photon,left,tension=0}{v1,v2}
\fmf{plain,tension=1.5}{v2,b}
\end{fmfgraph*}
\end{fmffile}}
\quad\ \ }
+\!\!\!
\subfigure[]{\parbox{5mm}{
\label{fig:wfgraph2}
\begin{fmffile}{wfgraph2}
\begin{fmfgraph*}(30,75)
\fmfset{arrow_len}{3mm}
\fmftop{t}
\fmfbottom{b}
\fmfv{l=$Q$,l.a=90,l.d=1.5mm}{t}
\fmfv{l=$Q^\dag$,l.a=-90,l.d=1.5mm}{b}
\fmf{plain,tension=1.5}{t,v1}
\fmf{fermion}{v1,v2}
\fmfset{wiggly_slope}{80}
\fmf{fermion,right,tension=0,width=.8}{v2,v1}
\fmf{photon,right,tension=0,width=.8}{v2,v1}
\fmf{plain,tension=1.5}{v2,b}
\end{fmfgraph*}
\end{fmffile}}\quad\ \ }
\qquad\qquad\quad
\subfigcapskip=5mm
\subfigure[]{\!\!\parbox{10mm}{
\label{fig:D_exchange_gauge}
\begin{fmffile}{D_exchange_gauge}
\begin{fmfgraph*}(30,90)
\fmfset{arrow_len}{3mm}
\fmftop{Qd1,Q1}
\fmfbottom{Q2,Qd2}
\fmfv{l=$\ \ {Q^\dag}^{ia}$,l.a=90}{Qd1}
\fmfv{l=$\ \ Q_{aj}$,l.a=90}{Q1}
\fmfv{l=$\ \ Q_{bk}$,l.a=-90}{Q2}
\fmfv{l=$\ \ {Q^\dag}^{lb}$,l.a=-90,l.d=.1w}{Qd2}
\fmf{plain}{Qd1,v1}
\fmf{plain}{v1,Q2}
\fmf{plain}{Q1,v2}
\fmf{plain}{v2,Qd2}
\fmfset{dot_len}{1.1mm}
\fmf{dots,tension=0,width=1.5}{v1,v2}
\end{fmfgraph*}
\end{fmffile}}}
\quad+\quad\,
\subfigure[]{\!\!\parbox{10mm}{
\label{fig:gluon_exchange_gauge}
\begin{fmffile}{gluon_exchange_gauge}
\begin{fmfgraph*}(30,90)
\fmfset{arrow_len}{3mm}
\fmftop{Qd1,Q1}
\fmfbottom{Q2,Qd2}
\fmfv{l=$\ \ {Q^\dag}^{ia}$,l.a=90}{Qd1}
\fmfv{l=$\ \ Q_{aj}$,l.a=90}{Q1}
\fmfv{l=$\ \ Q_{bk}$,l.a=-90}{Q2}
\fmfv{l=$\ \ {Q^\dag}^{lb}$,l.a=-90,l.d=.1w}{Qd2}
\fmf{plain}{Qd1,v1}
\fmf{plain}{v1,Q2}
\fmf{plain}{Q1,v2}
\fmf{plain}{v2,Qd2}
\fmfset{dot_len}{1.1mm}
\fmf{photon,tension=0}{v1,v2}
\end{fmfgraph*}
\end{fmffile}}}
\qquad\qquad\quad
\subfigure[]{\!\!\parbox{10mm}{
\label{fig:D_exchange_flavor}
\begin{fmffile}{D_exchange_flavor}
\begin{fmfgraph*}(30,90)
\fmfset{arrow_len}{3mm}
\fmftop{Q1,Qd1}
\fmfbottom{Qd2,Q2}
\fmfv{l=$\ \ {Q^\dag}^{ib}$,l.a=90}{Qd1}
\fmfv{l=$\ \ Q_{ai}$,l.a=90}{Q1}
\fmfv{l=$\ \ Q_{dj}$,l.a=-90}{Q2}
\fmfv{l=$\ \ {Q^\dag}^{cj}$,l.a=-90,l.d=.1w}{Qd2}
\fmf{plain,left=.3}{Qd1,v1}
\fmf{plain,left=.3}{v2,Q2}
\fmf{plain,right=.3}{Q1,v1}
\fmf{plain,right=.3}{v2,Qd2}
\fmfset{dot_len}{1.1mm}
\fmf{dots,tension=1.5,width=1.5}{v1,v2}
\end{fmfgraph*}
\end{fmffile}}}
\quad+\quad\,
\subfigure[]{\!\!\parbox{10mm}{
\label{fig:gluon_exchange_flavor}
\begin{fmffile}{gluon_exchange_flavor}
\begin{fmfgraph*}(30,90)
\fmfset{arrow_len}{3mm}
\fmftop{Q1,Qd1}
\fmfbottom{Qd2,Q2}
\fmfv{l=$\ \ {Q^\dag}^{ib}$,l.a=90}{Qd1}
\fmfv{l=$\ \ Q_{ai}$,l.a=90}{Q1}
\fmfv{l=$\ \ Q_{dj}$,l.a=-90}{Q2}
\fmfv{l=$\ \ {Q^\dag}^{cj}$,l.a=-90,l.d=.1w}{Qd2}
\fmf{plain,left=.3}{Qd1,v1}
\fmf{plain,left=.3}{v2,Q2}
\fmf{plain,right=.3}{Q1,v1}
\fmf{plain,right=.3}{v2,Qd2}
\fmf{photon,tension=1.5}{v1,v2}
\end{fmfgraph*}
\end{fmffile}}}
\label{fig:1loopgraphs}
\caption{1-loop Feynman graphs contributing to logarithmic divergences in two-point functions of generalized single-trace operators in the planar limit.  Dotted lines represent the auxiliary field $D^A$ in the vector multiplet.  All other relevant 1-loop graphs are obtained by replacing $Q,Q^\dag\leftrightarrow\tl Q^\dag,\tl Q$ on one or more scalar lines.}
\end{figure}

Returning to SQCD, we have the task of computing 1-loop corrections to two-point functions of bare operators $\cO_i^{(0)}$ and extracting $V$ --- the coefficient of their $1/\e$ poles.  In the planar limit, these corrections come from two sources.  The first is wavefunction renormalization for individual fields making up $\cO_i^{(0)}$, for example graphs \ref{fig:wfgraph1} and \ref{fig:wfgraph2}.  The sum of these graphs turns out to vanish in Feynman gauge (where the Euclidean gluon propagator is simply $g_{\mu\nu}/k^2$).  Hence we choose to use Feynman gauge in what follows.

The second contribution to $V$ is from interactions between pairs of neighboring fields in $\cO_i^{(0)}$.  Such pairs can either be gauge-contracted (e.g. graphs \ref{fig:D_exchange_gauge} and \ref{fig:gluon_exchange_gauge}) or flavor-contracted (e.g. graphs \ref{fig:D_exchange_flavor} and \ref{fig:gluon_exchange_flavor}).  For gauge-contracted pairs, we must normalize each graph by dividing by the tree-level two-point function $N_c\De_{xy}^2$, where
\be
\De_{xy}&\equiv&\frac{\G(1-\e)}{4\pi^{2-\e}}\frac{1}{|x-y|^{2-2\e}}
\ee
is the scalar propagator in $d=4-2\e$ dimensions.  We find, 
\be
\ref{fig:D_exchange_gauge}:&&
-\frac1{N_c\De_{xy}^2}\,
g^2\de_k^i\de^l_j \Tr(T^A T^A)\mu^{2\e}\!\!\int d^d z\,\De_{xz}^2\De_{zy}^2
\label{eq:D_exchange_gauge}
\ \ =\ \  -\de_k^i\de^l_j\frac{\l}{2\e}+\textrm{finite},\\
\ref{fig:gluon_exchange_gauge}:&&
\frac1{N_c\De_{xy}^2}\,
g^2\de_k^i\de^l_j \Tr(T^A T^A)\mu^{2\e}\!\!\int d^{d} z\,d^d w\,(\De_{xz}\lrptl^\mu_z \De_{zy})(g_{\mu\nu}\De_{zw})(\De_{xw}\lrptl_w^\nu\De_{wy})\nn\\
\label{eq:gluon_exchage_gauge}
&&=\ \ \de_k^i\de^l_j\frac{\l}{2\e}+\textrm{finite},
\ee
where $\l\equiv \frac{N_cg^2}{8\pi^2}$ is the 't Hooft coupling.

The divergent parts of Eqs.~(\ref{eq:D_exchange_gauge}) and (\ref{eq:gluon_exchage_gauge}) exactly cancel, so that nearest-neighbor interactions between gauge-contracted $Q^\dag Q$ pairs vanish at 1-loop.  However, if we replace one of the $Q$'s with $\tl Q^\dag$, then the $D$-term contribution \ref{fig:D_exchange_gauge} flips sign, while \ref{fig:gluon_exchange_gauge} is unchanged.  Consequently, nearest-neighbor interactions between gauge-contracted pairs take the following form at 1-loop:
\be
\label{eq:gaugecontractedcontribution}
V_\textrm{gauge contr's} &=& \left(
\begin{array}{c|cccc}
                  &Q^\dag Q & Q^\dag \tl Q^\dag & \tl Q Q & \tl Q \tl Q^\dag\\
                  \hline
Q Q^\dag          & 0    &                   &         &\\
Q \tl Q           &         &  -1             &         &\\
\tl Q^\dag Q^\dag &         &                   & -1    &\\
\tl Q^\dag \tl Q  &         &                   &         & 0\\
\end{array}
\right)
\x \l\,\id_\textrm{flavor} .
\ee

For interactions between flavor-contracted pairs, we must divide by the tree-level two-point function $N_f\De_{xy}^2$.  The graph \ref{fig:gluon_exchange_flavor} vanishes by antisymmetry of the gauge-scalar-scalar coupling.  The remaining contribution \ref{fig:D_exchange_flavor} is identical to \ref{fig:D_exchange_gauge}, up to overall factors.  Once again, the sign flips if we replace $Q\leftrightarrow \tl Q^\dag$, so we find the following contributions to $V$ from flavor-contracted neighbors:
\be
V_\textrm{flavor contr's} &=& \left(\begin{array}{c|cccc}
                  &QQ^\dag   & \tl Q^\dag\tl Q \\
                  \hline
Q^\dag Q          &  1        &-1 \\
\tl Q \tl Q^\dag  & -1       & 1 \\
\end{array}\right)
\x \frac{N_f}{2N_c}\l\x 2(T^A)_a^b (T^A)_d^c .
\ee
Whenever our flavor-contracted pair is part of a chain of more than two fields, the factor $2(T^A)_a^b (T^A)_c^d=\de_a^c\de^b_d-\frac 1 {N_c}\de_a^b\de_d^c$ is equivalent to the identity $(\id_\textrm{gauge})_{ad}^{cb}=\de_a^c\de^b_d$ in the planar limit.  However, there is an important exception for two-field operators, where each gauge generator is traced-over, giving $\Tr(T^A)\Tr(T^A)=0$.  In this case, the na\"ively subleading piece $-\frac 1 {N_c}\de_a^b\de_d^c$ contributes at the same order in $N_c$.

From here, it's easy to read off the dimensions of operators containing two fields.  Contributions from flavor contractions vanish, so that the 1-loop anomalous dimension matrix is given simply by Eq.~(\ref{eq:gaugecontractedcontribution}).  The mesons $\tl Q Q$ have 1-loop anomalous dimension $-\l$, consistent with their superconformal $R$-charge.  Meanwhile, both $Q^\dag Q$ and $\tl Q \tl Q^\dag$ have vanishing 1-loop anomalous dimensions, consistent with the fact that the baryon current $Q^\dag Q-\tl Q \tl Q^\dag$ is protected to all loops, while the Konishi operator $Q^\dag Q+\tl Q\tl Q^\dag$ has dimension equal to the slope of the $\b$-function at the conformal fixed point, which has its first nonzero contribution at two loops~\cite{Anselmi:1996mq}.

\subsection{Spin Chain Hamiltonian for General Operators}

To discuss more general operators, it is useful to introduce the notion of operators as spin chain states, with $V$ acting as a Hamiltonian.  Generalized single-trace operators are built out of strings of gauge adjoints $X=QQ^\dag$ and $Y=\tl Q^\dag \tl Q$, which we will call up states $|\!\!\up\>$ and down states $|\!\!\down\>$, respectively.  In this language, we may write
\be
\label{eq:gaugecontrspin}
V_\textrm{gauge contr's} &=& \frac \l 2(\s^z\otimes \s^z-\id\otimes\id),\\
\label{eq:flavorcontrspin}
V_\textrm{flavor contr's} &=& \frac{N_f}{2N_c}\l(\id-\s^x),
\ee
where $\s^x$ and $\s^z$ are the usual $2\x 2$ Pauli matrices.  Then traces $\Tr(XYX\dots)$ of length $L$ correspond to states $|\!\up\down\up\dots\>$ in a circular spin chain with $L$ sites.  For such states, the Hamiltonian is simply a sum of Eqs.~(\ref{eq:gaugecontrspin}) and (\ref{eq:flavorcontrspin}) for each site,
\be
\label{eq:closedchainhamiltonian}
V_\textrm{closed} &=& \sum_{n=0}^{L-1}\frac\l 2(\s_n^z\s_{n+1}^z-1)+\sum_{n=0}^{L-1}\frac{N_f}{2N_c}\l(1-\s_n^x)\nn\\
&=& \l\frac{N_f-N_c}{2N_c}L+\frac \l 2 \sum_{n=0}^{L-1}\p{\s_n^z\s_{n+1}^z-\frac{N_f}{N_c}\s_n^x},
\ee
where $\s^z_{L}\equiv\s_0^z$.  Note that this Hamiltonian commutes with the parity operator $P \equiv \prod_{n=0}^{L-1}\s_n^x$, and can be diagonalized separately on each of its eigenspaces $P=\pm 1$.

Similarly, we can think of the flavor adjoint and bi-fundamental operators in Eq. (\ref{eq:opensingletrace}) as states in an open spin chain, with nondynamical sites at each end.  For example, $Q^\dag XY\dots X Q$ and $\tl Q XY\dots X Q$ correspond to $|\!\!\up:\up\down\dots\up:\up\>$ and $|\!\!\down:\up\down\dots\up:\up\>$, respectively, where we have separated off the nondynamical sites with colons.  There is no flavor contraction at a nondynamical site, so the Hamiltonian for an open chain of length $L+2$ is
\be
V_\textrm{open} &=& \sum_{n=0}^{L}\frac\l 2(\s_n^z\s_{n+1}^z-1)+\sum_{n=1}^{L}\frac{N_f}{2N_c}\l(1-\s_n^x)\nn\\
&=&
\label{eq:openchainV}
\l\frac{N_f-N_c}{2N_c}L-\frac \l 2+\frac \l 2\p{\sum_{n=0}^{L}\s_n^z\s_{n+1}^z-\frac{N_f}{N_c}\sum_{n=1}^L\s_n^x} .
\ee

In the next section we will proceed to diagonalize Eqs.~(\ref{eq:closedchainhamiltonian}) and (\ref{eq:openchainV}) in order to obtain the 1-loop spectrum of operator dimensions.  Before we do so, let us mention that the generalization of these results to SQCD in the conformal window with an $\SO(N_c)$~\cite{Seiberg:1994pq,Intriligator:1995id} or $\Sp(N_c)$~\cite{Intriligator:1995ne} gauge group (in the Veneziano limit) is entirely straightforward.  In both cases, the only difference is that the states with odd parity are projected out (i.e., they vanish identically), but otherwise the Feynman diagrams and spectra of 1-loop operator dimensions are the same.

\section{Solving the Spin Chain}
\label{sec:solution}

\subsection{Closed Chains}

Eq.~(\ref{eq:closedchainhamiltonian}) is the Hamiltonian for a 1-dimensional anti-ferromagnetic Ising model in a transverse magnetic field.  Remarkably, this is a textbook example of an integrable model~\cite{Pfeuty197079,Sachdev}, and we will be able to write down its energy levels and eigenstates exactly.  The following solution is standard, but we include it here for completeness.

We will study the Hamiltonian
\be
H &=& \sum_{n=0}^{L-1} \left( \s_n^z\s_{n+1}^z-h\s_n^x \right) ,
\ee
for a magnetic field $h$ acting on a circular spin chain of length $L$.  Since our Hilbert space is a tensor product of two possible states for each site, it's tempting to think of it as a fermionic Fock space, with creation and annihilation operators $\s_n^{\pm}\equiv\frac 1 2(\s_n^y\pm i\s_n^z)$ for each position $n$.
This is not quite correct, since these operators commute rather than anticommute at different sites $\{\s_n^{\pm},\s^{\pm}_m\}\neq 0$.  The way around this problem is to perform a Jordan-Wigner transformation~\cite{JordanWigner}, defining fermionic creation and annihilation operators:
\be
\label{eq:JWtrans}
c_n^\dag \ \ \equiv\ \  \prod_{m=0}^{n-1}\s_m^x \s_n^+,
\qquad\qquad
c_n \ \ \equiv\ \  \prod_{m=0}^{n-1}\s_m^x \s_n^-,
\ee
which now satisfy canonical anticommutation relations
\be
\{c_n^\dag,c_m\} \ =\ \de_{nm},\qquad\{c_n,c_m\}\ =\ \{c_n^\dag,c_m^\dag\}\ =\ 0.
\ee
These new creation and annihilation operators are {\it nonlocal} on the spin chain, and we can think of them as creating and destroying fermionic solitons.

In these new variables, we have
\be
H &=& \sum_{n=0}^{L-2} (c_n^\dag+c_n)(c_{n+1}^\dag-c_{n+1})-P(c_{L-1}^\dag+c_{L-1})(c_{0}^\dag-c_{0})-h\sum_{n=0}^{L-1}(2c_n^\dag c_n -1)
\ee
where $P = \prod_{n=0}^{L-1}\s_n^x=\exp\p{i\pi\sum_n c_n^\dag c_n}=(-1)^F$ is the parity operator.  Within each parity eigenspace, the Hamiltonian becomes simply
\be
H &=& \sum_{n=0}^{L-1} (c_n^\dag+c_n)(c_{n+1}^\dag-c_{n+1})-h\sum_{n=0}^{L-1}(2c_n^\dag c_n -1),
\ee
where when $P=-1$ we must impose periodic boundary conditions $c_L=c_0$, and when $P=1$ we must impose antiperiodic boundary conditions $c_L=-c_0$.\footnote{Note that states with $P=1$ are exactly those created by an even number of $c$'s or $c^\dag$'s, so the states themselves will have periodic boundary conditions and can make sense on a circular chain.}  Our Hamiltonian is now translationally invariant and quadratic in creation and annihilation operators, so it can be diagonalized via a Fourier transform and Bogoliubov transformation:
\be
H &=& \sum_k \lr[{-(2\cos k+2h)c_k^\dag c_k-i\sin k\p{c^\dag_{-k}c_k^\dag+c_{-k}c_k}}]+Lh\\
&=& \sum_k \e(k)\p{b_k^\dag b_k-\frac 1 2}\label{eq:diagonalizedhamiltonian},
\ee
where $b_k^\dag$ and $b_k$ are new canonically normalized creation and annihilation operators, and
\be
\label{eq:dispersionrelation}
\e(k)&=& 2\sqrt{h^2+2h\cos(k)+1}
\ee
is the dispersion relation for the free fermionic quasiparticles created by $b^\dag_k$.  Because of the boundary conditions imposed by parity, the quasimomenta must take the values
\be
\label{eq:circularchainquantization}
k=\left\{
\begin{array}{cl}
\frac{2m\pi}{L} & P=-1\\
\frac{(2m+1)\pi}{L} & P=1
\end{array}\right.
\ee
for $m=0,1,\dots,L-1$.

\subsubsection{Examples: Closed Four-Field and Six-Field Operators}

Specializing to the case of interest, we can now write
\be
\label{eq:SQCDdiagonalizedhamiltonian}
V_\textrm{closed} &=& \l L + \l \sum_k \p{b_k^\dag b_k-\frac 1 2}\sqrt{10+6\cos(k)},
\ee
where we have used $h=\frac{N_f}{N_c}\aeq 3$ near the weakly-coupled end of the conformal window.  In addition to having quasimomenta obeying the quantization conditions in Eq.~(\ref{eq:circularchainquantization}), states corresponding to traces $\Tr(XYX\dots)$ must be invariant under shifts of the spin chain --- in other words they must have vanishing total momentum modulo $2\pi$.

As an example, let us consider four-field generalized single-trace operators, spanned by $\Tr(XX),\Tr(YY),$ and $\Tr(XY)$.  The two-site spin chain has two eigenstates of odd parity,
\be
b_{\pi}^\dag|0\>\qquad\textrm{and}\qquad b_{0}^\dag|0\>,
\ee
but only the latter has vanishing momentum.  It corresponds to the operator $\Tr(XX)-\Tr(YY)$, with anomalous dimension
\be
 \left( 2+\frac{1}{2}\sqrt{10+6\cos(0)}-\frac{1}{2}\sqrt{10+6\cos(\pi)} \right)\l &=& 3\l .
\ee
Meanwhile, we have two eigenstates of even parity, both of which have vanishing momentum,
\be
|0\>\qquad\textrm{and}\qquad b_{\frac{3\pi}2}^\dag b_{\frac \pi 2}^\dag|0\>.
\ee
These have anomalous dimensions
\be
 \left(2\pm\frac{1}{2}\sqrt{10+6\cos(\pi/2)}\pm\frac{1}{2}\sqrt{10+6\cos(3\pi/2)} \right)\l &=& (2 \pm \sqrt{10})\l,
\ee
and correspond to linear combinations of $\Tr(XY)$ and $\Tr(XX)+\Tr(YY)$.

As another simple example, let us repeat the above analysis for six-field generalized single-trace operators, built out of linear combinations of $\Tr(XXX)$, $\Tr(XXY)$, $\Tr(XYY)$, and $\Tr(YYY)$.   Using the $L=3$ quantization condition applied to the Hamiltonian Eq.~(\ref{eq:SQCDdiagonalizedhamiltonian}), the zero-momentum states for each parity and their associated anomalous dimensions are given by:
\be
\begin{array}{c|c|c}
P & \textrm{states} & \g \\
\hline
-1 & b_{0}^\dag|0\>,\ b_{4\pi/3}^\dag b_{2\pi/3}^\dag b_{0}^\dag|0\> & (5\pm \sqrt 7)\l\\
1 &  |0\>,\ b_{5\pi/3}^\dag b_{\pi/3}^\dag |0\> & (2 \pm \sqrt{13}) \l
\end{array}
\ee

It should be clear that the 1-loop spectrum of dimensions for closed operators with arbitrary values of $L$ can be similarly obtained.

\subsubsection{Asymptotics}
\label{sec:disc}

\begin{figure}
\begin{center}
\includegraphics{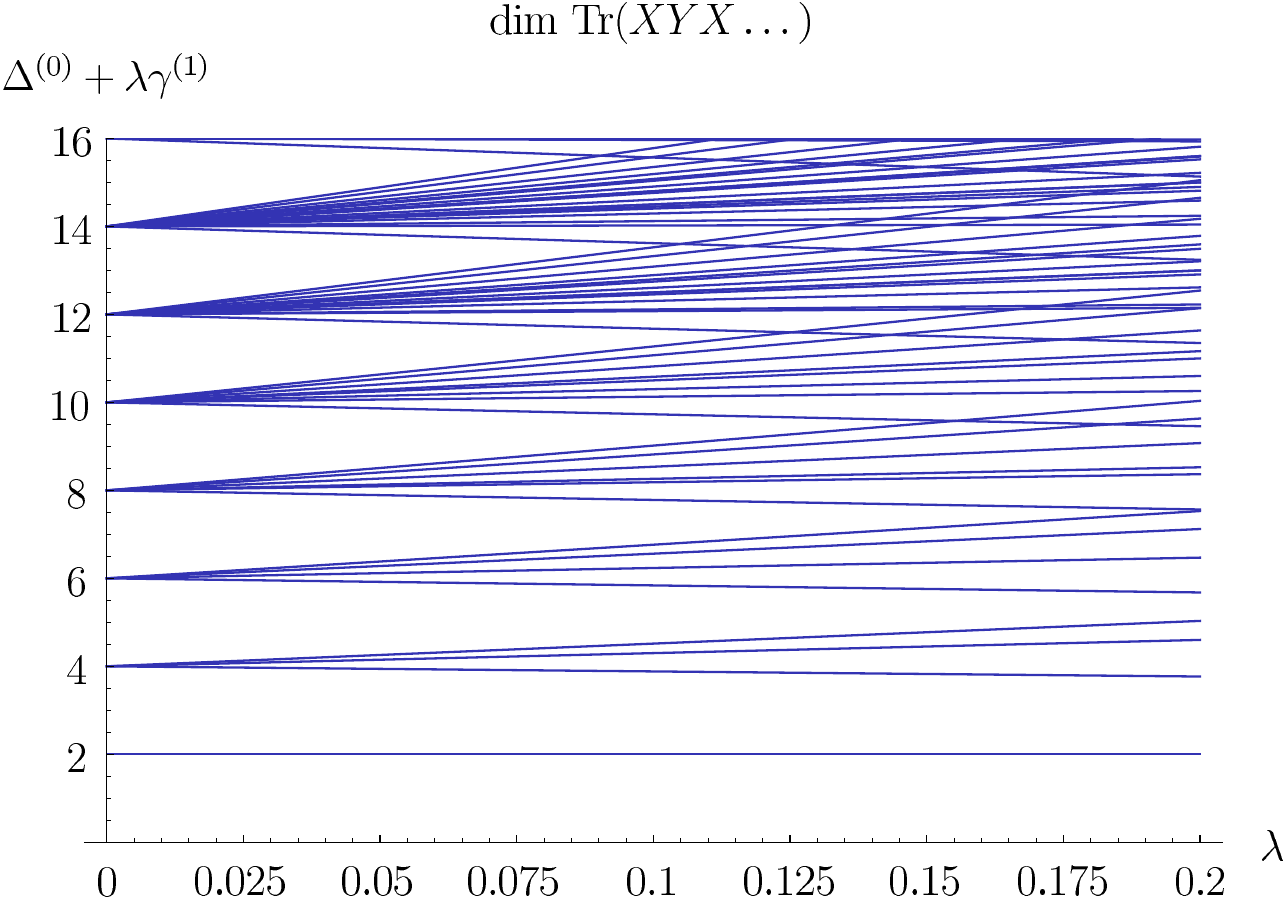}
\end{center}
\caption{Dimensions of flavor-singlet operators of the form $\Tr(XYX\dots)$ in $\cN=1$ SQCD in the Veneziano limit, as a function of $\l\simeq 3N_c/N_f-1$, up to 1-loop.}
\label{fig:levelcrossing}
\end{figure}

Armed with the explicit expression Eq.~(\ref{eq:SQCDdiagonalizedhamiltonian}), we can make some interesting observations about the pattern of operator dimensions for large classical dimension $\De^{(0)}=2L$.  In the limit $L\to\oo$, the distribution of momenta becomes continuous, and the highest and lowest anomalous dimensions $\g_{\pm}$ for a given $L$ can be approximated as elliptic integrals
\be
\g_{\pm} &=& \l L \pm \frac{\l L}{4\pi}\int_0^{2\pi}dk\sqrt{10+6\cos(k)}\nn\\
&=&\p{1\pm \frac{4E(3/4)}{\pi}} \l L\ \ \aeq\ \ (1\pm1.542)\,\l L.
\ee
Note that $|\g_{\pm}|$ becomes arbitrarily large as $L$ grows, so that dimensions of operators with different $L$ begin to cross.  Figure~\ref{fig:levelcrossing} shows the dimensions of generalized single trace operators of the form $\Tr(XYX\dots)$ for the first few $L$ and small values of $\l \simeq 3N_c/N_f-1$, and we can in fact see level crossing occur when $\l \gtrsim 1/(1.542L)$.

\subsection{Open Chains}

The case of open boundary conditions is slightly more complicated than that of a circular chain, but still exactly solvable at finite length.  The solution was first written down by Dou{\c c}ot et. al. in 2004 \cite{Doucot}, and clarified somewhat in \cite{Misquich}.  We will study the Hamiltonian
\be
\label{eq:openhamiltonian}
H &=& \sum_{n=0}^{L}\s_n^z\s_{n+1}^z-h\sum_{n=1}^L \s_n^x\nn\\
 &=& \sum_{n=0}^{L} (c_n^\dag+c_n)(c_{n+1}^\dag-c_{n+1})-h\sum_{n=1}^{L}(2c_n^\dag c_n -1),
\ee
acting on an open chain of length $L+2$.  Once again, we can diagonalize $H$ with an appropriate Fourier transform and Bogoliubov transformation.  However, we must take care with boundary conditions, since the spins at either end of the chain are nondynamical.  In particular, we should look for creation and annihilation operators that never mix eigenstates of $\s_0^z$ and $\s_{L+1}^z$.  Thus, let us consider the ansatz
\be
\label{eq:openchainansatz}
b_k^\dag &=& d_k^\dag - d_{-k}^\dag\\
d_k^\dag &=& \sum_{n=0}^{L+1}e^{ikn}(c_n^\dag-c_n)+f(k)e^{ikn}(c_n^\dag+c_n),
\ee
where $f(k)$ is to be determined, subject to the condition
\be
\label{eq:conditiononf}
f(k)e^{ik(L+1)}-f(-k)e^{-ik(L+1)} &=& 0.
\ee
Eq.~(\ref{eq:openchainansatz}) ensures that $\{b_k^\dag,\s_0^z\}=-i\{b_k^\dag,c_0^\dag-c_0\}=0$, while Eq.~(\ref{eq:conditiononf}) additionally ensures that $[b_k^\dag,\s_{L+1}^z]=0$.  In other words, $b_k^\dag$ flips $\s_0^z$ and preserves $\s_{L+1}^z$, never mixing different boundary spins.  Plugging our ansatz into the eigenvalue equation $[H,b_k^\dag]=\e(k)b_k^\dag$, we find
\be
-2f(k)(e^{-ik}+h) &=& \e(k)\\
-2(e^{ik}+h) &=& f(k)\e(k),
\ee
which has solution $\e(k)=2\sqrt{h^2+2h\cos(k)+1}$, as before.  The boundary condition Eq.~(\ref{eq:conditiononf}) becomes a quantization condition for the quasimomenta $k$,
\be
\label{eq:openquantizationcondition}
\frac{\sin(k(L+2))}{\sin(k(L+1))}&=&-h.
\ee

When $h>\frac{L+2}{L+1}$, as is always case for us, Eq.~(\ref{eq:openquantizationcondition}) has $L$ real solutions $k\in[0,\pi]$, along with an additional ``bound state" solution of the form $k=\pi+i\g$.  Together, the $b_{k}^\dag$ generate a Fock-space with dimension $2^{L+1}$, which exactly matches the number of states with fixed rightmost spin, say $\s_{L+1}^z=1$.  Thus, after suitably normalizing our creation and annihilation operators, the Hamiltonian on this subspace can be written in the form Eq.~(\ref{eq:diagonalizedhamiltonian}).  The Hamiltonian on the other subspace $\s^z_{L+1}=-1$ is conjugate by an overall parity transformation and has an identical spectrum.

Finally, let us identify which class of operators correspond to the highest and lowest energy states. When $h\ll 1$, it's clear that the highest energy state has ferromagnetic boundary conditions $\s_0^z=\s_{L+1}^z$.  One can show that no level-crossings occur as $h$ varies, so that this is the case for arbitrary $h>0$.  The ground state is obtained by acting with all $L+1$ lowering operators, and thus satisfies $\s_0^z|0\>=(-1)^{L+1}\s_{L+1}^z|0\>$.

\subsubsection{Example: Open Four-Field Operators}
Specializing now to SQCD, the open-chain Hamiltonian Eq.~(\ref{eq:openchainV}) can be written
\be
\label{eq:openanomalousdims}
V_\textrm{open} &=& \l \p{L-\frac 1 2}+\l \sum_k\p{b_k^\dag b_k-\frac 1 2}\sqrt{10+6\cos(k)},
\ee
with $k$ subject to Eq.~(\ref{eq:openquantizationcondition}).
As a simple example, let us consider open-chain operators with $L=1$.  Our quantization condition has two relevant solutions:
\be
k_1 =\cos^{-1}\p{\frac{-3+\sqrt{13}}{4}}\aeq 1.419,\qquad
k_2=\cos^{-1}\p{\frac{-3-\sqrt{13}}4}\aeq \pi-1.087 i.
\ee

Without loss of generality, we can consider states with positive rightmost spin.  The highest and lowest energy states both have ferromagnetic boundary conditions, so they will be linear combinations of $Q^\dag X Q$ and $Q^\dag Y Q$.  These correspond to the states
\be
|0\>\qquad\textrm{and}\qquad b_{k_1}^\dag b_{k_2}^\dag|0\>,
\ee
which have the anomalous dimensions
\be
\left(\frac{1}{2} \pm \frac12 \sqrt{10+6\cos(k_1)} \pm \frac12 \sqrt{10+6\cos(k_2)} \right) \l &=& \frac 1 2(1\pm \sqrt{13}) \l .
\ee
Linear combinations of $\tl QXQ$ and $\tl Q Y Q$ correspond to the remaining states
\be
b_{k_1}^\dag|0\>\qquad\textrm{and}\qquad b_{k_2}^\dag|0\>,
\ee
which then give the anomalous dimensions
\be
\left(\frac{1}{2} \pm \frac12 \sqrt{10+6\cos(k_1)} \mp \frac12 \sqrt{10+6\cos(k_2)} \right) \l &=& \left(\frac 1 2 \pm \frac 3 2\right) \l .
\ee
Dimensions of operators with open flavor indices at larger values of $L$ can be easily obtained in a similar fashion.

\section{Conclusions}
\label{sec:concl}

In the present work we have shown that the 1-loop dilatation operator of $\cN=1$ SQCD (in the electric Banks-Zaks limit) is equivalent to the 1-dimensional Ising spin chain in a transverse magnetic field, when acting on operators built out of scalar quarks.  There are a number of directions that merit further study.  One obvious direction would be to study the 1-loop dilatation operator in other sectors of the theory, as well as higher-order corrections.  In particular, one might ask whether the 2-loop dilatation operator in the scalar sector is solvable in any sense.  It would be interesting to see if, for instance, a nontrivial $S$-matrix for the Ising model fermions is induced at 2-loops.

A related direction would be to study the dilatation operator in the magnetic dual description of $\cN=1$ SQCD, which can be computed in perturbation theory when $N_f \sim \frac32 N_c$.  Because the dual description contains both dual quarks $q^{ai}, \tl{q}_{a\tl{\imath}}$ and mesons $M_{i}^{\tl{\imath}}$, generalized single-trace operators containing scalars may be built out of chains of the color-adjoint (flavor-singlet) objects $q (M M^\dag)^k q^\dag$, $\tl{q}^\dag (M^\dag M)^k \tl{q}$, $q (M M^\dag)^k M \tl{q}$, and $\tl{q}^\dag (M^\dag M)^k M^\dag q^\dag$.  Since there is an infinite tower of possible states at each site, it would be interesting to understand if the 1-loop dilatation operator in this basis can be described by a simple noncompact spin chain, as well as to investigate its integrability properties.

The transverse Ising model that we have encountered is one of the simplest examples of a spin chain with a quantum phase transition~\cite{Sachdev}.  The transition from a paramagnetic to a long-range ordered phase occurs at $h=1$, or na\"ively when $N_f/N_c = 1$, outside the conformal window and well outside the regime where perturbation theory is valid.  Nevertheless, it is tempting to speculate that this phase transition persists in the full theory, but gets lifted to the bottom of the conformal window $N_f/N_c = 3/2$.  Studying higher-order corrections and also the dual magnetic description should help to shed light on whether or not this is the correct picture.

Finally, there are many other 4D conformal field theories containing fundamental flavors that possess a Veneziano limit.  Studying these theories is important because they help us to break outside of the $\cN=4$ SYM universality class.  One particularly simple example is $\cN=2$ SQCD with $N_f = 2 N_c$ flavors, where the 1-loop dilatation operator acting on scalars shares many features with that of $\cN=1$ SQCD~\cite{Gadde:2009dj,Gadde:2010zi}.  Another example is non-supersymmetric QCD in the conformal window~\cite{Banks:1982} and its many variations.  It is an important goal of future research to better understand the extent to which these theories have solvable planar limits, and in particular to understand the role that supersymmetry plays in creating the kinds of simplifying features that have enabled so much progress in unraveling the structure of $\cN=4$ SYM.

\section*{Acknowledgements}

We would like to thank Clay C\'ordova, Sean Hartnoll, and Xi Yin for helpful comments and conversations.  This work is supported in part by the
Harvard Center for the Fundamental Laws of Nature and by NSF grant PHY-0556111.



\end{document}